# The Sun's Dynamic Extended Corona Observed in Extreme Ultraviolet

Daniel B. Seaton*[1,2], J. Marcus Hughes[1,2], Sivakumara K. Tadikonda[3], Amir Caspi[4], Craig DeForest[4], Alexander Krimchansky[5], Neal E. Hurlburt[6], Ralph Seguin[6], Gregory Slater[6]

[1] Cooperative Institute for Research in Environmental Sciences, University of Colorado, Boulder, Colorado
[2] National Centers for Environmental Information, National Oceanic and Atmospheric Administration, Boulder, Colorado
[3] Science Systems and Applications Inc., Lanham, Maryland
[4] Southwest Research Institute, Boulder, Colorado
[5] National Aeronautics and Space Administration, Goddard Space Flight Center, Greenbelt, Maryland
[6] Lockheed-Martin Solar and Astrophysics Laboratory, Palo Alto, California

**Abstract.** The "middle corona" is a critical transition between the highly disparate physical regimes of the lower and outer solar corona. Nonetheless, it remains poorly understood due to the difficulty of observing this faint region (1.5–3 $R_\odot$). New observations from the GOES Solar Ultraviolet Imager in August and September 2018 provide the first comprehensive look at this region's characteristics and long-term evolution in extreme ultraviolet (EUV). Our analysis shows that the dominant emission mechanism here is resonant scattering rather than collisional excitation, consistent with recent model predictions. Our observations highlight that solar wind structures in the heliosphere originate from complex dynamics manifesting in the middle corona that do not occur at lower heights. These data emphasize that low-coronal phenomena can be strongly influenced by inflows from above, not only by photospheric motion, a factor largely overlooked in current models of coronal evolution. This study reveals the full kinematic profile of the initiation of several coronal mass ejections, filling a crucial observational gap that has hindered understanding of the origins of solar eruptions. These new data uniquely demonstrate how EUV observations of the middle corona provide strong new constraints on models seeking to unify the corona and heliosphere.

## Our EUV Observations and the Middle Corona

The solar corona is the primary driver of almost all plasma dynamics throughout the solar system[1]. However, the precise nature of the connection between the corona and the heliosphere remains surprisingly poorly understood[2]. Recent solar and heliospheric observations taken by Parker Solar Probe, well within Mercury's orbit, revealed a highly structured environment shaped by flows and ejecta interacting with the corona's complex magnetic field[3,4,5,6]. The influence of these flows on the heliosphere and structural evolution on the global coronal topology is modulated by magnetic connectivity and physical transitions in the "middle corona" between 1.5 and 3 solar radii[7,8] ($R_\odot$; as measured from disk center). However, critical EUV observations of this region needed to characterize its plasma properties and dynamics have never before been made, and so significant gaps in our understanding persist. Even just predicting the appearance of the middle corona has been a major hurdle because[9] "as we are dealing with an unexplored region, *we really do not know how the fundamental plasma parameters vary with radial distance*. Even considering the simplest case of a streamer in the quiet Sun, very different estimates of densities and temperatures have been published." [Emphasis ours.]

---

*Correspondence to daniel.seaton@colorado.edu

Here we present the first ever long-duration observations of the middle corona in EUV. These observations directly address key questions about the characteristics of the middle corona and how processes within it influence the global structure of the corona-heliosphere system[10,11,12]. This includes the initiation and driving of eruptive solar flares and the acceleration of the solar wind. We discuss how these observations can provide much-needed constraints for theory and models[9,13,14,15] of the corona on global scales. Such models provide the only way to connect measured observables with the physical processes at their origins because comprehensive direct measurements of the corona are infeasible. However, models that use the observed photospheric magnetic field as their primary boundary condition are known to systematically underestimate the resulting interplanetary magnetic flux compared to *in situ* measurements[14]. Because changes in the polar surface fields manifest primarily as changes in structure of the middle corona[15], our new observations promise to provide valuable constraints for these global models.

The lack of direct measurements of the middle corona — particularly in the EUV — stems largely from historical biases, technical limitations, and optimization choices, but not from the physics of this region itself. EUV telescopes have focused primarily on the low corona, only to heights below 1.7 $R_\odot$, because of a commonly held misperception that EUV emission in the middle corona was too dim to observe with practical instrumentation[16,17,18]. Recently, limited EUV imaging studies in a single wavelength with PROBA2/SWAP[19,20,21] extended these observations somewhat, revealing unexpected structure and transient dynamics to about 2 $R_\odot$. However, no single EUV observation prior to our campaign covered the full spatial extent of the middle corona or provided temperature diagnostics over sufficiently long timescales to systematically characterize dynamics in this region.

Our new high-contrast EUV images from a 2018 deep-exposure wide-field mosaic campaign[22] of the Solar Ultraviolet Imager[23,24] on NOAA's GOES-17 spacecraft settle the debate over whether EUV emission is detectable in the middle corona, revealing structure and plasma dynamics throughout the region. Combining our SUVI data with LASCO[25] images at greater heights reveals complex relationships among phenomena in the lower corona, the evolution of coronal morphology on global scales, and the origins of discrete structure in the solar wind. This highlights, in particular, how outflows into the solar wind are frequently born in the middle corona and how feedback from the outer corona drives changes near the solar surface. This campaign demonstrates that deep-field EUV imaging provides a direct, experimental probe into coronal properties and dynamics that have previously been only indirectly inferred[26]. These observations open a window into a generally overlooked region of the Sun-Earth system, resolving over a half-century of uncertainty[27,28,29,30] about the magnetic and dynamic connection between the surface of the Sun and the solar wind that fills the heliosphere.

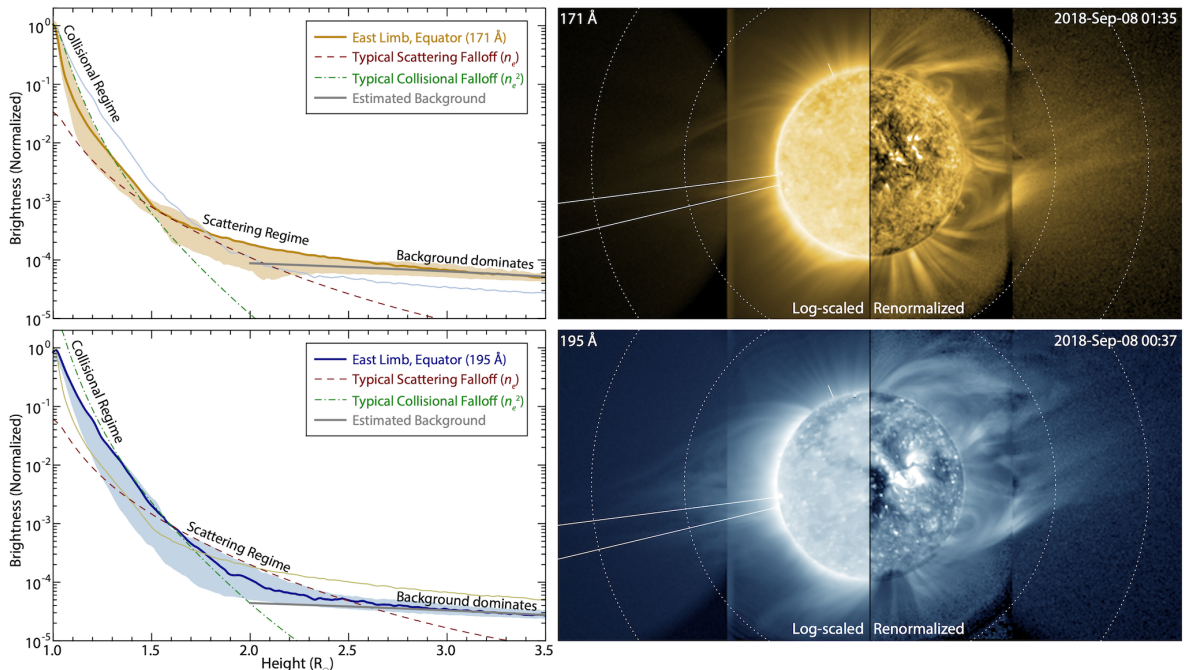

**Figure 1.** (Left) Normalized brightness falloff as a function of height along a radial cut of ~6° through a streamer. 171 Å and 195 Å curves are shown in both panels for reference. Error bars are smaller than the line thickness. The shaded region shows the span of streamer brightnesses between 10% and 90% of the minimum and maximum of across the entire east limb. Dashed lines show characteristic falloff expected from resonant scattering and collisional excitation, proportional to electron density and density squared, respectively. Overlays are measured for each respective wavelength individually. (Right) Comparison of natural (log-scaled) and renormalized views of the corona, indicating the cut region. The solar north pole is indicated by a small tick mark.

Because SUVI was not designed for off-pointed observation, careful processing was required to address the unique technical challenges and artifacts present in these data (see

Methods). One primary consideration was compensating for the intrinsic dynamic range of the observations ($10^4$–$10^6$ between the brightest active regions and the faintest detectable structures) using a normalizing radial function extracted from the data. This processing preserves the local contrast required to see coherent structures and motions by nearly eliminating the overall decrease in brightness with altitude (Figure 1).

The intensity of the EUV emission decreases by more than four orders of magnitude from the limb out to about 3.5 $R_\odot$ (Figure 1). The solid lines show the rate of falloff along the radial slice of a streamer indicated in the right-hand panels, selected because it appears prominently in both passbands. The shaded region shows the falloff considered along the entire east-limb streamer structure, over the 10-90 percentile range of all brightnesses at a given radial distance. This helps to visualize large scale trends in the falloff within in this coronal feature, bounding its behavior at global scale and across passbands.

The rate of falloff is diagnostic of the EUV emission mechanism. Dashed lines in Figure 1 show typical rates of falloff (see Golub & Pasachoff, Figure 1.4[31]) for emission due to collisional excitation (proportional electron density squared) and resonant or Thomson scattering (linearly proportional to electron density). Comparing the falloff along a radial slice of a streamer suggests that emission from this structure is consistent with thermally-driven collisional excitation below approximately 1.5 $R_\odot$. Larger-scale features above 1.5 $R_\odot$ are consistent with resonant scattering of emission originating elsewhere in the corona. Although this resonantly scattered emission is intrinsically faint, our processing overcomes this obstacle to enable visualization of coherent structure out to heights of 2–3 $R_\odot$, where the magnetic field becomes primarily radial and connected to the interplanetary magnetic field. Our study confirms recent simulations[13] that demonstrated that the fraction of resonantly scattered emission increases steeply at heights above 1.5 $R_\odot$, including in the Fe emission lines observed in the 195 Å SUVI passband. Further, these simulations highlight that observations of the ratio of resonant scattering to collisional excitation emission can serve as a novel probe of the previously unmeasured solar wind speed in this region.

**Dynamics**

Our observations reveal three major classes of dynamic processes that continuously reshape the middle corona that have not been systematically characterized before: (1) gradual bulk motion that disrupts and reshapes the global coronal structure, manifesting as inflows and outflows of a few km/s traceable to heights of at least 4 $R_\odot$; (2) small-scale outflows with velocities of roughly 50–150 km/s that originate in cooler, plume-like features close to the solar surface and are visible to 2–3 $R_\odot$; and (3) impulsive eruptions that are linked to CMEs observed by LASCO, which rapidly alter coronal topology on global scales. Coronagraphic studies of the outer corona[32] above 5 $R_\odot$ detected largely uniform inflow and outflow with small but significant variations embedded within. This campaign, providing the first continuous single-instrument observations connecting the lower and middle corona, reveals that motion in the middle corona is much less homogenous, suggesting a strong interdependence between structure, dynamics, and the highly localized physical properties of their regions of origin.

The use of two simultaneous passbands, 171 and 195 Å, provides the first extended temperature diagnostics in the middle corona to >3 R$_\odot$. Figure 1 highlights the differences in observable structures between these two passbands. In the collisionally dominated regime, below ~1.5 R$_\odot$, emission in these passbands is generally from ions with characteristic temperatures around 0.8 and 1.5 MK, respectively[33]. At larger heights this emission reflects only ionization state, which is influenced by, but no longer directly coupled to, temperature. This ionization selectivity becomes clearly apparent in Figure 2 and the accompanying animation, where the two passbands are overlain in false color. For example, the large loop structure off the northwest limb (Figure 2 D; also visible in the separate co-temporal images in Figure 1) is readily identifiable as a multi-thermal structure, potentially with complex 3D geometry, in the composite image. The movie reveals that this feature forms and retracts over a period of 12 to 18 hours, and its complex thermal structure is likely the result of reconnection-associated heating during the loop formation. In contrast, the polar plumes are cooler features predominantly visible in the 171 Å image, illustrating the difference in energetics between phenomena in closed versus open magnetic structures in the middle corona.

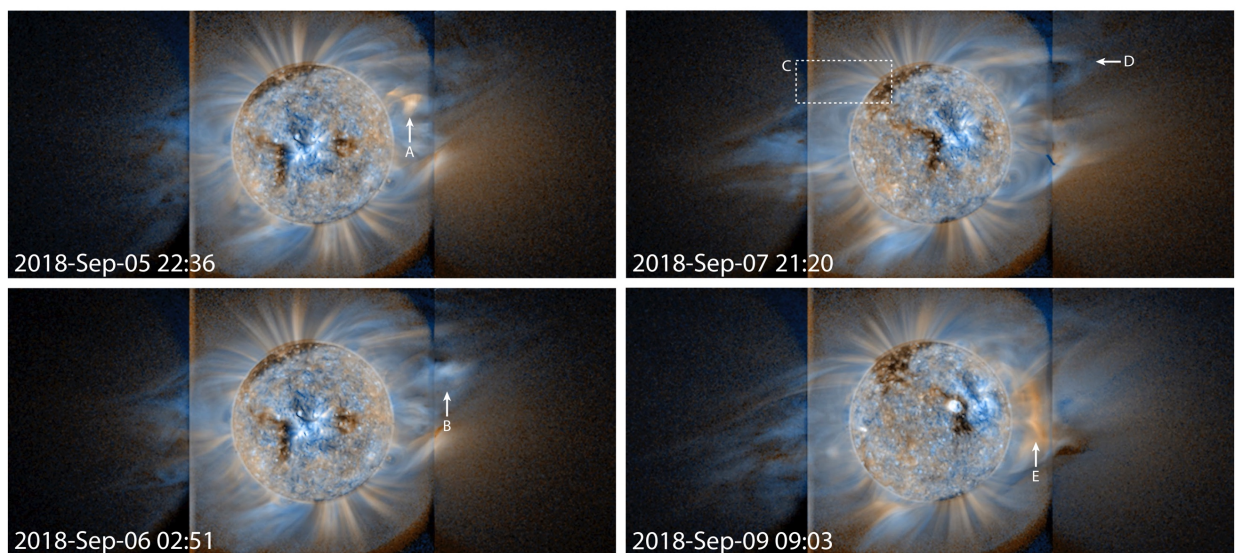

**Figure 2.** Selected composite SUVI images at 171 Å (gold) and 195 Å (blue). Features delineated by the arrows and box are discussed in the text and correspond to the same labels in Figures 3 and 4. These labels highlight features of interest in the accompanying animations in the supplementary materials. All images have the same orientation as indicated in Figure 1. Supplementary Table 1 provides a complete overview of these features.

Combining these new wide-field EUV images with visible-light coronagraphy from LASCO (Figure 3 and accompanying animations) reveals how the flows that SUVI observes from their inception are transformed by their passage through the turbulent and complex middle corona, establishing the outflows and structure observed in the outer corona and heliosphere[32,34,35,36]. As these flows propagate outward, the plasma β – the ratio of the gas kinetic pressure to the magnetic pressure – changes from low (magnetic field dominates) to high (gas dynamics dominate). It is not known exactly where this transition occurs, nor how it influences the flow evolution.

It has often been assumed this β transition solely occurs high in the corona, and in open structures such as polar plumes, where plasma can flow freely outwards, this β transition likely does occur beyond this campaign's field of view[37] (FOV). In closed structures such as loops, however, gas pressure can accumulate and can overwhelm the magnetic field. This allows fluid instabilities to drive runaway processes including large-scale outward and inward flows that drag the frozen-in magnetic field along with them, feeding back to drive evolution and changes in magnetic topology even in the middle and low corona, within our FOV. The dynamics observed by SUVI strongly suggest that this transition occurs even below 2 $R_\odot$ in closed structures, most obviously along the streamer belt[10] (Figure 3 F, G, & H and co-temporal animation frames).

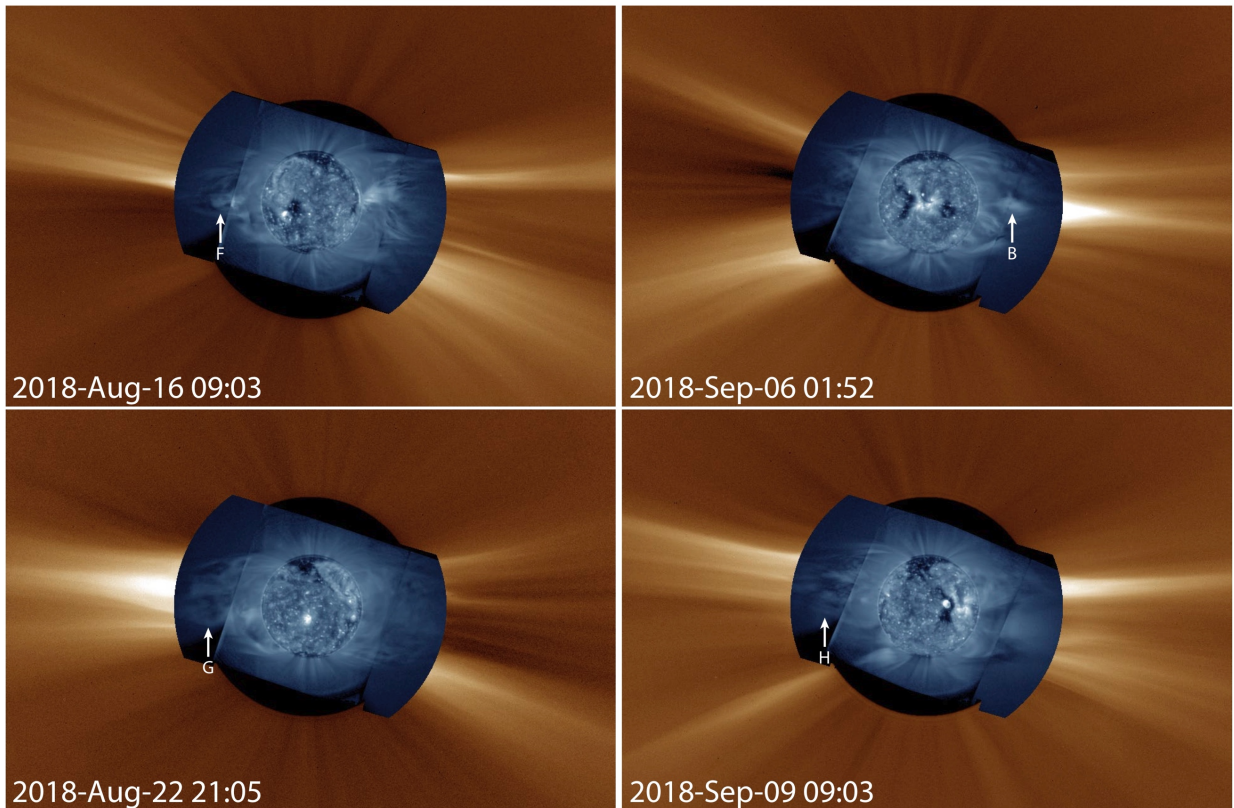

**Figure 3.** Selected SUVI 195 Å and WL LASCO C2 coronagraph composite images. Features delineated by the arrows are discussed in the text and correspond to the same labels in Figures 2 and 4. These labels highlight features of interest in the accompanying animations in the supplementary materials. The SUVI image is terminated at 2.7 $R_\odot$, the LASCO image extends to about 6 $R_\odot$ in the east-west direction and 4 $R_\odot$ in the north-south direction. Images are oriented with solar-north directly upwards. Supplementary Table 1 provides a complete overview of these features.

We visualize the temporal evolution of the three types of observed flows using height-time diagrams to track feature intensity along selected radial slices above the limb (Figure 4).

***Gradual Bulk In- and Outflows:*** Figure 4 F & G show the dynamic evolution of the bulk inflow and outflow features highlighted by Figure 3 F & G. This type of flow is characterized by relatively gradual evolution and velocities of only a few km/s and is continuously traceable from SUVI into the surrounding LASCO FOV. In SUVI, these flows are observed primarily in the warmer 195 Å channel, indicating a close relationship between plasma heating and the underlying flow acceleration process. Their emergence from closed-field regions requires reconfiguration of the underlying magnetic field[38], a process that liberates stored magnetic energy and heats the surrounding corona[39]. SUVI's month-long campaign demonstrates how long-term systematic observations allow us to characterize the coupling between these flows and the likely role of closed-field reconnection in constantly reshaping the global morphology and connectivity of the corona across this region.

Long-term tracking of these flows could help address the "missing heliospheric flux problem"[14,15] by providing constraints on the significance of time-varying processes in determining this magnetic connectivity.

Gradual flows that originate higher in the corona than impulsive eruptions have also been suggested to be a potential source of so-called "stealth CMEs," that is, CMEs that are visible in coronagraphs but lack clear low-coronal sources[40,41]. In general, the velocities of such events are also slower than typical CMEs[42]. Without EUV observations of the middle corona available for past studies, the sources of outflows that originate in this region could only very rarely be identified directly, but our new observations close that gap. Indeed, several of the gradual outflows observed during this campaign (e.g. Figure 3 G & H) may be candidate stealth CME sources.

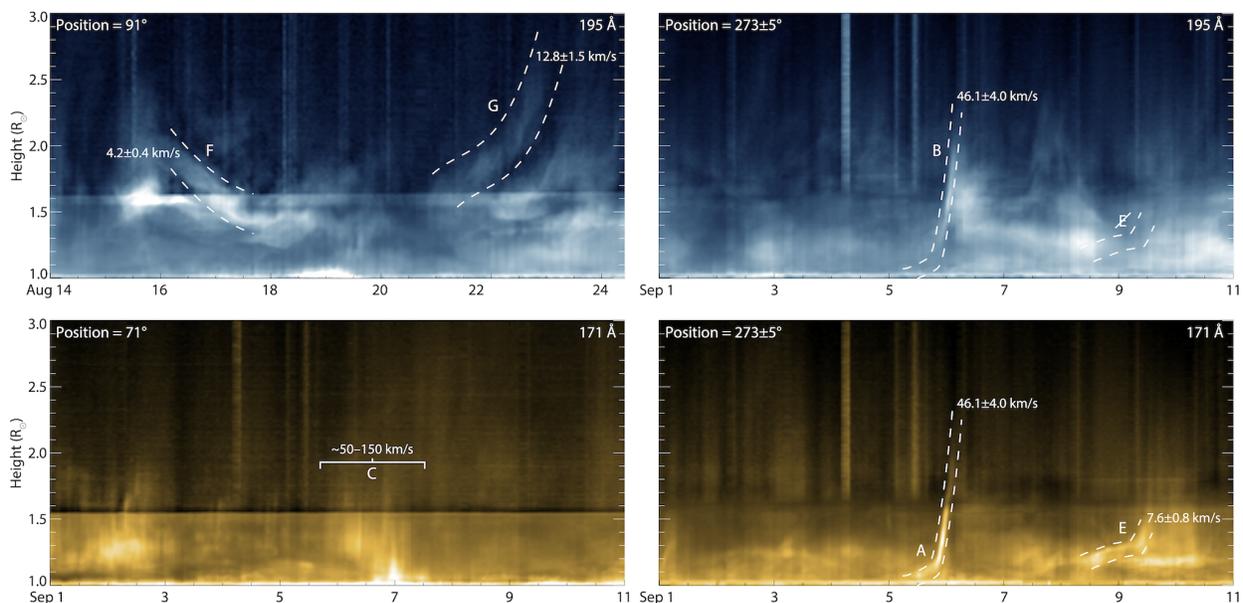

**Figure 4.** Height-time diagrams showing radial evolution of features as a function of time at selected positions, wavelengths, and time ranges. Dashed lines highlight coherent flows for visual reference. Velocities are indicated near the location where they were measured. Features delineated by letters are discussed in the text and correspond to the same labels in Figures 2 and 3.

Flows that begin in the middle corona and propagate *inwards*, such as those highlighted in Figure 3 & Figure 4 F, are particularly interesting. These flows can disturb and even destabilize existing structures when they arrive near the solar surface. Inflows in the outer corona have been extensively cataloged[47], but these flows have not generally been observed originating in or propagating through the middle corona outside of specific events linked directly to solar eruptions. Thus, it has generally been assumed that low-coronal dynamics are driven from below, i.e., that the dominant driver of the low corona is emergence and subsequent motion of magnetic field at the solar surface. Our observations reveal that the middle corona is a significant, important influence on the low corona, as these inflows drive changes from *above* — an interaction that has gone largely untreated in the literature. In this case, the inflow observed in SUVI is associated with corresponding motions in LASCO. This may be an EUV manifestation of a "streamer detachment"[48], demonstrating the role of the middle corona in mediating feedback on the low corona by events initiated at much higher altitudes. These observations emphasize the need for further dedicated study and modelling of this mechanism driving low-coronal dynamics.

***Small-scale rapid outflows:*** In contrast to the above, flows that occur in open field regions (Figure 2 and Figure 4 C) propagate unimpeded by overlying magnetic forces and therefore have relatively little influence on global coronal dynamics. They are rapid and cool, and have nearly constant profiles as they travel outwards. Such small-scale jets are apparently ubiquitous in the low corona[43,44,45] and have been associated with narrow, intermittent outflows observed by coronagraphs[46]. Our new SUVI observations directly track these features through the middle corona and highlight how future extended studies can quantify the role such flows play in carrying energy, structure, and magnetic flux from the low corona, where they are initiated, into the solar wind throughout the heliosphere.

***Impulsive Eruptions:*** Our campaign occurred near the nadir of solar minimum, when the global field is very nearly dipolar and little free magnetic energy is available to drive eruptions. Nonetheless, we observed several modest CMEs (Figure 2 A, B, & E; Figure 3 B; Figure 4 A, B, & E) that reached speeds of a few tens of km/s. Two of these CMEs (Figure 2 and Figure 4 A, B, & E) highlight how these observations permit us to identify the precise location of impulsive acceleration and infer the temperatures throughout the events. The faster CME on Sep 6 (e.g., Figure 4 A & B) clearly warms during acceleration, visible as a dimming in (cooler) 171 Å and simultaneous brightening in (hotter) 195 Å. (Although the 171 Å dimming occurs at the seam between the sun-pointed and off-pointed images in the mosaic, contemporaneous observations from PROBA2/SWAP and the sun-pointed SUVI on GOES-16 confirm the dimming trend begins in the low corona and is not an instrumental artifact.)

In contrast, the slow, weak CME on Sep 9 stays cool throughout (Figure 4 E), with no noticeable emission in 195 Å (Figure 4; note the region marked $\not\!E$), an indication — coupled with its slower velocity — that this event is less energetic overall. The height at which impulsive acceleration begins, visible as the knee from small to large slope in the height-time profile, also differs between these two events. The differences between these two representative examples occurring in essentially the same location on the Sun — and between others in the dataset, not shown here — illustrate the clear need for systematic studies that can sufficiently sample the broad range of physical conditions under which CMEs manifest and the processes that initiate them.

Although eruption initiation has been extensively modeled, specific predictions about CME acceleration and magnetic reconnection-driven energy release have not been sufficiently validated due to the lack of appropriate observations in the middle corona. Many models[49,50,51] predict that CME acceleration occurs primarily below 2 $R_\odot$. Only a handful of observations[52,53,55] have captured complete CME trajectories that track their acceleration through this region, or the associated impulsive expansion of the CME as it rises[54]. Likewise, models of reconnection in eruptions predict the formation of current sheets that drive flows and heating in the middle corona, but these have been observed only infrequently[23,56,57,58] and often the reconnection site itself lies outside the FOV of low coronal imagers[59]. While the few, weak CMEs we observed in this campaign are not optimal to directly address these questions, they prove that dedicated wide-field, multi-wavelength EUV observations during periods of greater activity can extend our existing capabilities[23,54] to provide the necessary constraints to fully validate CME initiation models. Such observations would also greatly benefit space weather forecasting by helping to refine the predictions of propagation speeds and potential geoeffectiveness of major CMEs.

**Discussion**

Globally, space agencies have recently prioritized missions to observe and connect the outer corona and heliosphere through their dynamic interface, with both remote sensing and *in situ* measurements. These include current missions Parker Solar Probe and Solar Orbiter, the upcoming Polarimeter to Unify the Corona and Heliosphere (PUNCH) and Aditya-L1, along with new missions dedicated to space weather operations at multiple vantage points in the heliosphere. New ground-based coronagraphs that can probe the coronal magnetic field through spectropolarimetry, to heights of about 2 $R_\odot$, are also in development[60]. Though these missions are expected to provide groundbreaking insight into the physics of their respective domains, a comprehensive understanding of the intertwined corona-heliosphere system as a whole requires observations that can capture the pivotal physics in the middle corona.

Our observations address long-persistent questions about the transition from the magnetically dominated low corona to the gas-dynamic dominated outer corona. They strongly indicate that the middle corona indeed encompasses the transition from low to high $\beta$ in closed structures, as well as a topological shift from magnetic loops to predominantly radial structure. This campaign reveals directly how this region mediates the outward flow that gives rise to the solar wind and its embedded magnetic field in addition to the multidirectional flow of information as the global corona adjusts to perturbations at particular locations. The observations demonstrate that EUV emission in the middle corona, whether collisionally excited or from resonant scattering, can provide valuable diagnostics of the temperature, density, connectivity and dynamics of this region.

Direct observations of the time-dependent evolution of the middle corona such as these provide important new constraints on — and validation of — model results, and can help reconcile inconsistencies between existing predictions and *in situ* measurements of the conditions in the outer corona and heliosphere. Our exploratory, long-duration, deep-field SUVI campaign, the first of its kind, captured a variety of phenomena that demonstrate the complexity and importance of this region, even during solar minimum. This pathfinding campaign, and a follow-on in late 2019 using improved procedures, lay the foundation for future observations dedicated to high-sensitivity, long-term, wide-FOV EUV observations that will lead to a holistic understanding of the Sun's entire outer atmosphere.

## References

1. Schwenn, R. Space Weather: The Solar Perspective. *Living Rev. Sol. Phys.* **3**, 2. https://doi.org/10.12942/lrsp-2006-2 (2006).
2. Cranmer, S. R., & Winebarger, A. R. The Properties of the Solar Corona and Its Connection to the Solar Wind. *Ann. Rev. Astron. Astrophys.* **57**, 157–187. https://doi.org/10.1146/annurev-astro-091918-104416 (2019).
3. Howard, R. A., et al. Near-Sun observations of an F-corona decrease and K-corona fine structure. *Nature* **576**, 232–236. https://doi.org/10.1038/s41586-019-1807-x (2019).
4. Kasper, J. C., et al. Alfvénic velocity spikes and rotational flows in the near-Sun solar wind. *Nature* **576**, 228–231. https://doi.org/10.1038/s41586-019-1813-z (2019).
5. McComas, D. J., et al. Probing the energetic particle environment near the Sun. *Nature* **576**, 223–227. https://doi.org/10.1038/s41586-019-1811-1 (2019).
6. Bale, S. D., et al. Highly structured slow solar wind emerging from an equatorial coronal hole. *Nature* **576**, 237–242. https://doi.org/10.1038/s41586-019-1818-7 (2019).
7. DeForest, C. E., Howard, R. A., Velli, M., Viall, N., & Vourlidas, A. The Highly Structured Outer Solar Corona. *Astrophys. J.* **862**, 18. https://doi.org/10.3847/1538-4357/aac8e3 (2018).
8. Chhiber, R., Usmanov, A. V., Matthaeus, W. H., & Goldstein, M. L. Contextual Predictions for the Parker Solar Probe. I. Critical Surfaces and Regions. *Astrophys. J. Suppl. Ser.* **241**, 11. https://doi.org/10.3847/1538-4365/ab0652 (2019).
9. Del Zanna, G., Raymond, J., Andretta, V., Telloni, D., & Golub, L. Predicting the COSIE-C Signal from the Outer Corona up to 3 Solar Radii. *Astrophys. J.* **865**, 132. https://doi.org/10.3847/1538-4357/aadcf1 (2018).
10. Vásquez, A. M., van Ballegooijen, A. A., & Raymond, J. C. The Effect of Proton Temperature Anisotropy on the Solar Minimum Corona and Wind. *Astrophys. J.* **598**, 1361–1374. https://doi.org/10.1086/379008 (2003).
11. Masson, S., McCauley, P., Golub, L., Reeves, K. K., & DeLuca, E. E. Dynamics of the Transition Corona. *Astrophys. J.* **787**, 145. https://doi.org/10.1088/0004-637X/787/2/145 (2014).
12. DeForest, C. E., Hassler, D. M., & Schwadron, N. A. On the Magnetic Correspondence between the Photosphere and the Heliosphere. *Sol. Phys.* **229**, 161–174. https://doi.org/10.1007/s11207-005-5376-9 (2005).
13. Gilly, C. R. & Cranmer, S. R. The Effect of Solar Wind Expansion and Nonequilibrium Ionization on the Broadening of Coronal Emission Lines. *Astrophys. J.* **901**, 150. https://doi.org/10.3847/1538-4357/abb1ad (2020).
14. Linker, J. A., et al. The Open Flux Problem. *Astrophys. J.*, **848**, 70. https://doi.org/10.3847/1538-4357/aa8a70 (2017).
15. Riley, P., et al. Can an Unobserved Concentration of Magnetic Flux Above the Poles of the Sun Resolve the Open Flux Problem? **884**, 18. https://doi.org/10.3847/1538-4357/ab3a98 (2020).
16. Schrijver, C. J., & McMullen, R. A. A Case for Resonant Scattering in the Quiet Solar Corona in Extreme-Ultraviolet Lines with High Oscillator Strengths. *Astrophys. J.* **531**, 1121–1128. https://doi.org/10.1086/308497 (2000).

17. Winebarger, A. R., Warren, H. P., & Mariska, J. T. Transition Region and Coronal Explorer and Soft X-Ray Telescope Active Region Loop Observations: Comparisons with Static Solutions of the Hydrodynamic Equations. *Astrophys. J.* **587**, 439–449. https://doi.org/10.1086/368017 (2003).
18. DeForest, C. E., Martens, P. C. H., & Wills-Davey, M. J. Solar Coronal Structure and Stray Light in TRACE. *Astrophys. J.* **690**, 1264–1271. https://doi.org/10.1088/0004-637X/690/2/1264 (2009).
19. Seaton, D. B., De Groof, A., Shearer, P., Berghmans, D., & Nicula, B. SWAP Observations of the Long-term, Large-scale Evolution of the Extreme-ultraviolet Solar Corona. *Astrophys. J.* **777**, 72. https://doi.org/10.1088/0004-637X/777/1/72 (2013).
20. Goryaev, F., Slemzin, V., Vainshtein, L., & Williams, D. R. Study of Extreme-ultraviolet Emission and Properties of a Coronal Streamer from PROBA2/SWAP, Hinode/EIS and Mauna Loa Mk4 Observations. *Astrophys. J.* **781**, 100. https://doi.org/10.1088/0004-637X/781/2/100 (2014).
21. O'Hara, J. P., Mierla, M., Podladchikova, O., D'Huys, E., & West, M. J. Exceptional Extended Field-of-view Observations by PROBA2/SWAP on 2017 April 1 and 3. *Astrophys. J.* **883**, 59. https://doi.org/10.3847/1538-4357/ab3b08 (2019).
22. Tadikonda, S. K., et al. Coronal Imaging with the Solar UltraViolet Imager. *Sol. Phys.* **294**, 28. https://doi.org/10.1007/s11207-019-1411-0 (2019).
23. Seaton, D. B., & Darnel, J. M. Observations of an Eruptive Solar Flare in the Extended EUV Solar Corona. *Astrophys. J.* **852**, L9. https://doi.org/10.3847/2041-8213/aaa28e (2018).
24. Vasudevan, G., et al. Design and on-orbit calibration of the solar ultraviolet imager (SUVI) on the GOES-R series weather satellite. *Proc. SPIE* **11180**, 111807P. https://doi.org/10.1117/12.2536196 (2019).
25. Brueckner, G. E., et al. The Large Angle Spectroscopic Coronagraph (LASCO). *Sol. Phys.* **162**, 357–402. https://doi.org/10.1007/BF00733434 (1995).
26. Strachan, L., et al. Latitudinal dependence of outflow velocities from O VI Doppler dimming observations during the Whole Sun Month. *J. Geophys. Res.* **105**, 2345–2356. https://doi.org/10.1029/1999JA900459 (2000).
27. Parker, E. N. Dynamics of the Interplanetary Gas and Magnetic Fields. *Astrophys. J.* **128**, 664. https://doi.org/10.1086/146579 (1958).
28. Leighton, R. B., Noyes, R. W., & Simon, G. W. Velocity Fields in the Solar Atmosphere. I. Preliminary Report. *Astrophys. J.* **135**, 474. https://doi.org/10.1086/147285 (1962).
29. Hundhausen, A. J. *Physics and Chemistry in Space 5: Coronal Expansion and Solar Wind* (ed. Roederer, J. G.) (Springer: Berlin, 1990).
30. Cranmer, S. R. Coronal Holes and the High-Speed Solar Wind. *Space Sci. Rev.* **101**, 229–294. https://doi.org/10.1023/A:1020840004535 (2002).
31. Golub, L. & Pasachoff, J. M. *The Solar Corona (2nd ed.)*. (Cambridge University Press., 2009)
32. DeForest, C. E., Howard, T. A., & McComas, D. J. Inbound Waves in the Solar Corona: A Direct Indicator of Alfvén Surface Location. *Astrophys. J.* **787**, 124. https://doi.org/10.1088/0004-637X/787/2/124 (2014).

33. O'Dwyer, B., Del Zanna, G., Mason, H. E., Weber, M. A., & Tripathi, D. SDO/AIA response to coronal hole, quiet Sun, active region, and flare plasma. *Astron. Astrophys.* **521**, A21. https://doi.org/10.1051/0004-6361/201014872 (2010).
34. Sheeley, N. R., et al. Measurements of Flow Speeds in the Corona Between 2 and 30 $R_\odot$. *Astrophys. J.* **484**, 472–478. https://doi.org/10.1086/304338 (1997).
35. Viall, N. M., Spence, H. E., Vourlidas, A., & Howard, R. A. Examining Periodic Solar-Wind Density Structures Observed in the SECCHI Heliospheric Imagers. *Sol. Phys.* **267**, 175–202. https://doi.org/10.1007/s11207-010-9633-1 (2010).
36. Rouillard, A. P., et al. The Solar Origin of Small Interplanetary Transients. *Astrophys. J.* **734**, 7. https://doi.org/10.1088/0004-637X/734/1/7 (2011).
37. DeForest, C. E., Matthaeus, W. H., Viall, N. M., & Cranmer, S. R. Fading Coronal Structure and the Onset of Turbulence in the Young Solar Wind. *Astrophys. J.* **828**, 66. https://doi.org/10.3847/0004-637X/828/2/66 (2016).
38. Pontin, D. I., & Wyper, P. F. The Effect of Reconnection on the Structure of the Sun's Open-Closed Flux Boundary. *Astrophys. J.* **805**, 39. https://doi.org/10.1088/0004-637X/805/1/39 (2015).
39. Seaton, D. B., & Forbes, T. G. An Analytical Model for Reconnection Outflow Jets Including Thermal Conduction. *Astrophys. J.* **701**, 348–359. https://doi.org/10.1088/0004-637X/701/1/348 (2009).
40. Robbrecht, E., Patsourakos, S., & Vourlidas, A. No Trace Left Behind: STEREO Observation of a Coronal Mass Ejection Without Low Coronal Signatures. *Astrophys. J.* **701**, 283–291. http://dx.doi.org/10.1088/0004-637X/701/1/283 (2009).
41. Ma, S., Attrill, G. D. R, Golub, L. & Lin, J. Statistical Study of Coronal Mass Ejections With and Without Distinct Low Coronal Signatures. *Astrophys. J.* **722**, 289–301. http://dx.doi.org/10.1088/0004-637X/722/1/289 (2010).
42. D'Huys, E., Seaton, D. B., Poedts, S., & Berghmans, D. Observational Characteristics of Coronal Mass Ejections Without Low-Coronal Signatures. *Astrophys. J.* **795**, 49. http://dx.doi.org/10.1088/0004-637X/795/1/49 (2014).
43. Dobrzycka, D., Cranmer, S. R., Raymond, J. C., Biesecker, D. A., & Gurman, J. B. Polar Coronal Jets at Solar Minimum. *Astrophys. J.* **565**, 621–629. https://doi.org/10.1086/324431 (2002).
44. Savcheva, A., et al. A Study of Polar Jet Parameters Based on Hinode XRT Observations. *Publ. Astron. Soc. Jpn.* **59**, S771. https://doi.org/10.1093/pasj/59.sp3.S771 (2007).
45. Raouafi, N. E., et al. Solar Coronal Jets: Observations, Theory, and Modeling. *Space Sci. Rev.* **201**, 1–53. https://doi.org/10.1007/s11214-016-0260-5 (2016).
46. Wang, Y.-M., & Sheeley, N. R. Coronal White-Light Jets near Sunspot Maximum. *Astrophys. J.* **575**, 542–552. https://doi.org/10.1086/341145 (2002).
47. Sheeley, N. R. & Wang, Y.-M. Coronal Inflows During the Interval 1996–2014. *Astrophys. J.* **797**, 10. http://dx.doi.org/10.1088/0004-637X/797/1/10 (2014).
48. Sheeley, N. R. & Wang, Y.-M. In/Out Pairs and the Detachment Of Coronal Streamers. Astrophys. J. 655, 1142–1156. https://doi.org/10.1086/510323 (2007).

49. Chen, P. F. Coronal Mass Ejections: Models and Their Observational Basis. *Living Rev. Sol. Phys.* **8**, 1. https://doi.org/10.12942/lrsp-2011-1 (2011).
50. Lin. J. et al. Review on Current Sheets in CME Development: Theories and Observations. *Space Sci. Rev.* **194**, 237–302. https://doi.org/10.1007/s11214-015-0209-0 (2015).
51. Chen, J., & Krall, J. Acceleration of coronal mass ejections. *J. Geophys. Res. (Space Phys.)* **108**, 1410. https://doi.org/10.1029/2003JA009849 (2003).
52. Bein, B. M., et al. Impulsive Acceleration of Coronal Mass Ejections. I. Statistics and Coronal Mass Ejection Source Region Characteristics. *Astrophys. J.* **738**, 191. https://doi.org/10.1088/0004-637X/738/2/191 (2011).
53. Bein, B. M., Berkebile-Stoiser, S., Veronig, A. M., Temmer, M., & Vršnak, B. Impulsive Acceleration of Coronal Mass Ejections. II. Relation to Soft X-Ray Flares and Filament Eruptions. *Astrophys. J.* **755**, 44. https://doi.org/10.1088/0004-637X/755/1/44 (2012).
54. Veronig, A., et al. Genesis and Impulsive Evolution of the 2017 September 10 Coronal Mass Ejection. *Astrophys J.* **868**, 107. https://doi.org/10.3847/1538-4357/aaeac5 (2018).
55. D'Huys, E., Seaton, D. B., De Groof, A., Berghmans, D., & Poedts, S. Solar signatures and eruption mechanism of the August 14, 2010 coronal mass ejection (CME). *J. Space Weath. Space Clim.* **7**, A7. https://doi.org/10.1051/swsc/2017006 (2017).
56. Savage, S. L., McKenzie, D. E., Reeves, K. K., Forbes, T. G., & Longcope, D. W. Reconnection Outflows and Current Sheet Observed with Hinode/XRT in the 2008 April 9 "Cartwheel CME" Flare. *Astrophys. J.* **722**, 329–342. https://doi.org/10.1088/0004-637X/722/1/329 (2010).
57. Warren, H. P., et al. Spectroscopic Observations of Current Sheet Formation and Evolution. *Astrophys. J.* **854**, 122. https://doi.org/10.3847/1538-4357/aaa9b8 (2018).
58. Longcope, D., Unverferth, J., Klein, C., McCarthy, M., & Priest, E. Evidence for Downflows in the Narrow Plasma Sheet of 2017 September 10 and Their Significance for Flare Reconnection. *Astrophys. J.* **868**, 148. https://doi.org/10.3847/1538-4357/aaeac4 (2018).
59. Seaton, D. B., Bartz, A. E., & Darnel, J. M. Observations of the Formation, Development, and Structure of a Current Sheet in an Eruptive Solar Flare. *Astrophys. J.* **835**, 139. https://doi.org/10.3847/1538-4357/835/2/139 (2017).
60. Landi, E., Habbal, S. R., & Tomczyk, S. Coronal plasma diagnostics from ground based observations. *J. Geophys. Res.* **121**, 8237–8249. https://doi.org/10.1002/2016JA022598 (2016).

## Methods

SUVI is a solar EUV telescope with six wavelength channels producing 1280×1280 pixel images with a 2.5 arcsec pixel scale, primarily intended for space weather operations[61]. For this campaign we used SUVI in an east-west rastering mode to construct three-panel mosaic images in three passbands, 171, 195, and 304 Å from 2018 August 7 to September 13. Only the first two channels yielded sufficient statistics for analysis. We used 10-s full-resolution exposures for the Sun-pointed central panel and 20-s 2×2 rebinned exposures for each of the side panels, at each wavelength. These images were normalized by exposure time, co-aligned on the ground, and assembled into complete images of the EUV corona that extend to approximately 5 $R_\odot$ in the horizontal direction. The full observation sequence, including all three passbands and repointing between offset positions, required a little over six minutes per cycle. Our observations coincided with the GOES eclipse season, meaning that brief data gaps appear around local midnight (about 05:30 UT) throughout the campaign.

SUVI is on the Sun-pointing platform mounted on the solar panel on the GOES spacecraft. Because the GOES spacecraft makes terrestrial observations from geostationary orbit, the solar array must rotate normal to Earth's equatorial plane to maintain a constant view to the Sun. As such, SUVI's horizontal imaging axis is parallel to the Earth's equatorial plane, which is generally not parallel to the solar equatorial plane. In our campaign images, the horizontal axis is rotated ~16–23° clockwise from the solar equator, depending on relative orientations of Earth and Sun.

We improve signal-to-noise in the Sun-pointed central field by rebinning that image 2×2 on the ground, resulting in a uniform platescale of 5 arcsec per pixel across the full mosaic. To further improve the signal-to-noise ratio in these mosaics, we median-stack sets of three composite images in time, yielding an effective imaging cadence of just under 20 minutes for the observations used in this analysis.

SUVI was intended to be operated with the Sun at the center of the FOV. Consequently, when used in off-pointed mode, the internal telescope baffles do not fully mitigate off-axis light, and a strong glint contaminated the off-pointed panels throughout this campaign. This glint is largely invariant on short timescales and is superimposed on the nominal solar signal. To remove this stray light, we computed a running boxcar minimum for each pixel with a time window of ±16 hours around each frame and subtracted it after smoothing to suppress artifacts.

The relatively high flux of low energy electrons in geostationary orbit contaminates most SUVI observations, producing spikes that are especially problematic in the long-duration side-panel exposures. A despiking step before image assembly suppresses the strongest transient noise resulting from these particles, but is ineffective in removing weak events. In the off-pointed panels, these weak events, coupled with photon counting (shot) noise, can be the same order as the true solar signal, and another strategy is required. Therefore, we further mitigate transient noise by applying a three-dimensional ($x \times y \times t$) Savitzky-Golay filter which works by locally fitting polynomials to small data windows tiled over the entire data cube[62,63,64]. This polynomial serves as a locally tunable smoothing function that suppresses weak transient noise. Lower-degree polynomials preserve only the coarse structure in the data while suppressing high frequency variation, while higher-degree polynomials preserve more fine detail at the risk of preserving more noise. For these data we used a 9×9×9 pixel window with a 5×5×5 degree polynomial, which provided the best balance between removing noise and preserving fine structure based on visual inspection of the data.

A final processing step is required to equalize the dynamic range for optimal display, as discussed in the main text (Figure 1). To do this, we generate an azimuthally varying radial filter, i.e., a polar-coordinate normalizing function that samples the radial falloff and smooths the azimuthal intensity over a 25° window, derived uniquely for each individual frame. The filter is then apodized to avoid over-enhancing noise at the radial extrema and over-suppressing brightness at the limb, which helps to preserve a more natural appearance. Each image is then renormalized by its unique filter. This is similar to—but less sophisticated than—the well-known Fourier Normalizing-Radial-Graded Filter[65]. This filtering process can, occasionally, generate a small ring-like artifact in regions of the limb where there is limited brightness (e.g. polar coronal holes) and such an artifact does appear in the movies and figures on occasion. Additional details on this method will be described in a forthcoming paper (DBS, in prep).

This processing compensates for the four-orders-of-magnitude falloff in coronal brightness globally, while preserving the intrinsic luminosity on local scales to retain coherent spatial features. All of the SUVI images and movies in this paper make use of the same background subtraction, noise reduction, and radial filter techniques except where noted in Figure 1. To enhance the visibility of features in the two-color display in Figure 2 we further increase the contrast using an unsharp mask; this processing is not required in the simpler single-channel SUVI component of Figure 3.

The LASCO data presented in Figure 3 were prepared using the standard LASCO data reduction tools distributed in the Interactive Data Language (IDL) by the LASCO team. To improve signal-to-noise in these images, we stacked them in time to a one-hour effective cadence. We removed stray light and F-corona from the stacked images by computing a minimum-value image from the entire campaign and subtracting it from each frame. An additional despiking step suppressed both cosmic rays and stars in the image to isolate the desired K-corona signal. We applied a single azimuthally isotropic and temporally invariant radial filter to each frame, derived from a median image calculated over the entire campaign, to equalize dynamic range similarly to SUVI.

## Methods References

61. Seaton, D. B. et al. Solar Dynamics. in *The GOES-R Series* (eds. Goodman, S., et al.) (Elsevier: Dordrecht). 219–232. https://doi.org/10.1016/B978-0-12-814327-8.00018-4 (2020).

62. Savitzky, A. & Golay, M. J. E. Smoothing and Differentiation of Data by Simplified Least Squares Procedures. *Anal. Chem.* **36**, 1627–1639. https://doi.org/10.1021/ac60214a047 (1964).
63. Shekhar, C. On Simplified Application of Multidimensional Savitzky-Golay Filters and Differentiators. *AIP Conf. Proc.* **1705**, 020014. https://doi.org/10.1063/1.4940262 (2016).
64. Leifer Lab. Github repository of Savitzky-Golay implementation. https://github.com/leiferlab/savitzkygolay (2018).
65. Druckmüllerová, H., Morgan, H., & Habbal, S. Enhancing Coronal Structures with The Fourier Normalizing-Radial-Graded Filter. *Astrophys. J.* **737**, 88. http://dx.doi.org/10.1088/0004-637X/737/2/88 (2011).
66. NOAA SUVI Data Archive & Metadata. https://doi.org/10.7289/V5FT8J93 (2020).

## Data Availability

Standard SUVI observations are available for download via the NOAA National Centers for Environmental Information GOES-R archive[66]. Preliminary data products from this campaign, as used in this paper, will be made available by request to the SUVI team at NCEI by email to goesr.suvi@noaa.gov or directly to the corresponding author. Fully processed SUVI observations will be made public at the same site as soon as possible, after final data product development for this campaign is complete and covid-related restrictions permit local access to the necessary network infrastructure.

## Code Availability

The data processing and analysis discussed in this paper leveraged publicly available software packages in *Python* and *SolarSoft IDL*. The specific processing steps that generated figures and movies presented in this paper used an iterative process that spanned several platforms and multiple languages, so publishing the code in a single, self-contained processing package is not straightforward. However, all processing codes will be provided on request.

## Acknowledgments

We acknowledge our colleagues at the GOES-R Program Office, the Lockheed-Martin SUVI team, and NOAA's National Centers for Environmental Information and Space Weather Prediction Center for providing support and helpful input during the development of this campaign and the analysis of results, particularly Pamela C. Sullivan, Gustave J. Comeyne, Margaret Shaw-Lecert, Robin R. Minor, Rob Redmon, Janet Machol, and Steven Hill. We thank the anonymous referees and journal editorial staff for thorough and thoughtful suggestions that improved the clarity of our text and figures and strengthened the presentation of our scientific conclusions. D.B.S. and J.M.H. acknowledge support for CIRES's GOES-R activities via NOAA cooperative agreement NA17OAR4320101; DBS acknowledges support from NASA Grant 80NSSC20K1283. A.C. acknowledges funding from NASA Grant NNX15AQ68G; A.C. & C.E.D. acknowledge NASA PUNCH Contract 80GSFC18C0014.

## Author Contributions

D.B.S. led the data analysis, image processing, and visualization efforts. J.M.H. developed image processing software. S.K.T. and A.K. developed the campaign and assisted with its implementation, data acquisition, and analysis. A.C. and C.E.D. assisted with interpretation of the data and development of data visualizations. N.E.H., R.S. and G.S. developed software for SUVI data calibration and assembly of mosaics and assisted with interpretation of data. D.B.S. and A.C. led the writing of the manuscript, with contributions from the other authors.

## Competing Interests

None.

# Supplementary Materials

*Note:* Because animations cannot be attached to directly to this arXiv preprint, please contact the corresponding author to request access to animations.

**Animation 1.** Composite SUVI movie at 171 Å (gold) and 195 Å (blue). The labeled panels in the corresponding Figure 2 highlight dynamic features and events that are of particular interest. Timestamps should permit the identification of these events within the movie. [see Figure_2_Animation.mp4]

**Animation 2.** Two-Panel rendering of Animation 1, at 171 Å (gold) and 195 Å (blue). [see Figure_2_Supplemental_2Panel.mp4]

**Animation 3.** SUVI 195 Å and visible-light LASCO C2 coronagraph composite movie. The labeled panels in the corresponding Figure 3 highlight dynamic features and events that are of particular interest. Timestamps should permit the identification of these events within the movie. [see Figure_3_Animation.mp4]

**Animation 4.** Uncropped rendering of SUVI 195 Å frames from Animation 3. Images are presented in natural SUVI camera coordinates, with celestial north oriented upwards and solar north rotated roughly 20° counterclockwise from vertical. [see Figure_3_Supplemental_SUVI_Only.mp4]

**Table 1.** Reference guide to features of interest highlighted in the figures and movies.

| Feature | Approximate location | Start Date & Time | Passband | Animation |
|---|---|---|---|---|
| A & B | West | Sep 05 15:00 | 171 & 195 | 2 & 3 |
| C | Northeast | Sep 06 00:00 | 171 | 2 |
| D | Northwest | Sep 07 20:00 | 171 & 195 | 2 |
| E | Southwest | Sep 09 04:00 | 171 & 195 | 2 & 3 |
| F | East | Aug 16 00:00 | 195 | 3 |
| G | East | Aug 22 00:00 | 195 | 3 |
| H | East | Sep 08 15:00 | 195 | 3 |